# Engineering of the qubit initialization in an imperfect physical system

**Tianfeng Chen[1,2], Lin Wan[1,2], Jiamin Qiu[1,2], Hong Peng[1,2], Jie Lu[3,4*], and Ying Yan[1,2*]**

[1]School of Optoelectronic Science and Engineering & Collaborative Innovation Center of Suzhou

Nano Science and Technology, Soochow University, Suzhou 215006, China

[2]Key Lab of Advanced Optical Manufacturing Technologies of Jiangsu Province & Key Lab of

Modern Optical Technologies of Education Ministry of China,

Soochow University, 215006 Suzhou, China

[3]Department of Physics, Shanghai University, 200444 Shanghai, China

[4]Shanghai Key Lab for Astrophysics, 100 Guilin Road, 200234 Shanghai, China

E-mail: lujie@shu.edu.cn (J. Lu) and yingyan@suda.edu.cn (Y. Yan).

**Abstract**

We propose a method to engineer the light matter interaction while initializing a qubit present of physical constraints utilizing the inverse engineering. Combining the multiple degrees of freedom in the pulse parameters with the perturbation theory, we develop pulses to initialize the qubit within a tightly packed frequency interval to an arbitrary superposition state with high fidelity. Importantly, the initialization induces low off-resonant excitations to the neighboring qubits, and it is robust against the spatial inhomogeneity in the laser intensity. We apply the method to the ensemble rare-earth ions system, and simulations show that the initialization is more robust against the variations in laser intensity than the previous pulses, and reduces the time that ions spend in the intermediate excited state by a factor of 17. The method is applicable to any systems addressed in frequency such as nitrogen-vacancy centers, superconducting qubits, quantum dots, and molecular qubit systems.

## 1. Introduction

High precision coherent control of quantum systems is essential to the state initialization and manipulation in quantum computing [1], quantum communication[2, 3], quantum simulation [4], and quantum metrology [5]. Among any complicated coherent controls for qubits, the startup step is the same, i.e. initializing a qubit to an arbitrary superposition state in fast timescale and with high fidelity. It is a challenge to achieve the high-fidelity initialization in some imperfect experimental systems, where unavoidable physical constraints are present. For example, the addressing frequency of an ensemble qubit in the rare-earth ions (REI) qubit system inhomogeneously spreads within a few hundred kHz [6, 7]. High fidelity manipulation on this ensemble qubit demands a uniform coherent control over this tightly packed frequency interval [8, 9]. Another constraint impeding the high-fidelity control is the unwanted off-resonant excitations, typically lying about a few MHz away from the qubit frequency, originating from other transitions of the neighboring qubits closely spaced in frequency [10-12]. Besides these two constraints, the third constraint in experiments which degrades the fidelity is the uneven spatial distribution in the laser intensity under detection. In most cases a plane wave with constant intensity is considered in theoretical works, which is true only for the very central part of a Gaussian beam. For this, one can use a pinhole in the detection system to intentionally select the central part of the beam for detection, but pays the price with the huge drop in the signal strength. Thus, it results in a low signal to noise ratio (SNR), therefore affects the accuracy of the measurements. High fidelity initialization on a qubit in such an imperfect system would require that the quantum control simultaneously overcome all these three constraints. Firstly, it is robust against the frequency detuning within a tightly packed frequency interval. Secondly, it causes negligible unwanted off-resonant excitations to the qubits close by in frequency. Thirdly, it is robust against the inhomogeneous variations in the laser intensity, or equivalently in the Rabi frequency. Our work aims for designing fast and robust pulses which are capable of initializing a qubit in such an imperfect physical system to an arbitrary superposition state.

To increase the robustness of the quantum control against various fluctuations (e.g., the noise in the amplitude or phase of the light source), variations (e.g., inhomogeneity in the qubit addressing frequency or in the intensity of a Gaussian beam) or inaccuracies (e.g., errors in time or other control parameters) in an imperfect system, two scenarios are often concerned. One is the quantum feedback control. the other is the open-loop control. A nice review of the quantum feedback can be found in this reference [13]. Quantum feedback is a closed-loop control, where a signal obtained from a quantum system is fed back to the system, and the controller acts on the system by adjusting one or more parameters based on the feedback signal until the predefined objective (e.g. robustness) is achieved. Although closed-loop control has clear advantage with respect to the abrupt or unexpected error sources present in the system, open-loop control is more effective in the face of the well-defined imperfections, such as variations in the qubit addressing frequency or inhomogeneous intensity distribution in a Gaussian beam. A number of open-loop control protocols have been proposed to enhance the robustness of the quantum control, for instance, the optimal control theory [14-16], Stark-chirped rapid adiabatic passage [17], sampling-based learning control [18], learning-based open-loop control [19], accelerated dissipation-based approach [20], and other protocols based on the shortcuts to adiabaticity (STA) [21-26]. Some of the open-loop control are adiabatic passages [27, 28], which is relatively slow limited by the adiabatic theorem. This potentially increases the decoherence. Here we focus on the nonadiabatic open-loop control, more specifically, the STA technique, which combines fast speed with high robustness, providing a nonadiabatic route towards the results which the adiabatic passage can achieve. There several techniques were investigated: inverse engineering based on the Lewis-Riesenfeld (LR) invariant [29-32], counter diabatic driving [22, 33], and fast-forward approach [34]. Each technique has different characteristics and may apply to different experimental systems [32]. Inverse engineering offers more flexibility to design the exact dynamical evolution while ensuring the target states via boundary conditions. It has been used to develop pulses which has high robustness against the variations in frequency, and at the same time causes negligible off-resonant excitations by optimizing the multiple degrees of freedom introduced in the pulse parameters [10]. However, to demonstrate the high-fidelity manipulation in experiments, a pinhole is needed in the detection system to intentionally select the most central part of the Gaussian beam where Rabi frequency varies less than 5% [35]. In this work we treat the variation in Rabi frequency as a perturbation, employ the perturbation theory to investigate the reduction in fidelity, and develop pulses which simultaneously overcome all three aforementioned constraints. We apply the method to the REI system, which is an ideal system to verify the constraints of both the frequency inhomogeneity and unwanted off-resonant excitations. Importantly, REI system is an excellent test bed for both quantum computation and quantum memory because of their long coherence properties [36]. A comparison with the pulses developed previously shows that the pulses in this work have better performance with respect to the Rabi frequency variations. The method is applicable to any imperfect systems addressed in frequency such as nitrogen-vacancy centers [37], superconducting qubits [11], quantum dots [38], and molecular qubit systems [11].

The article is arranged as follows. We first introduce the inverse engineering scheme in a three-level system and the systemic error sensitivity with the help of perturbation theory. Then we propose a method to construct the fast, robust, and low off-resonant excitation pulses, and show the performance of the pulses in simulation in application to the REI system. In the end we discuss the results and present the conclusions.

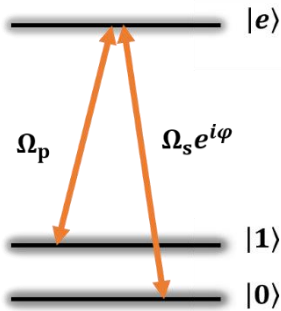

Figure 1. (Color online) Schematic energy-level diagram in a $\Lambda$ configuration. The qubit is represented by level $|0\rangle$ and $|1\rangle$, which are coupled by the optical transitions $|1\rangle - |e\rangle$ and $|0\rangle - |e\rangle$ with driving fields $\Omega_p$ and $\Omega_s e^{i\varphi}$, respectively, where $\varphi$ is a time-independent phase.

## 2. Invariant-based inverse engineering and the systematic error sensitivity

Under the rotating wave approximation, the Hamiltonian of a three-level system as shown in figure 1, reads [39]

$$H_0(t) = \frac{\hbar}{2}\begin{bmatrix} 0 & \Omega_p(t) & 0 \\ \Omega_p(t) & 0 & \Omega_s(t)e^{-i\varphi} \\ 0 & \Omega_s(t)e^{i\varphi} & 0 \end{bmatrix}, \quad (1)$$

where $\Omega_i$ $(i = s, p)$ is the Rabi frequency describing the coupling between light pulses and atomic transitions, defined as $\Omega_i = -\frac{\vec{\mu}_i \cdot \vec{E}_i}{\hbar}$ $(i = s, p)$. $\vec{\mu}_i$ denotes the transition dipole moments, $\vec{E}_i$ the electric field, and $\varphi$ a time-independent phase of the field $\Omega_s$. LR theory tells that the eigenstates of the LR invariants of the time-dependent Hamiltonian can be used to construct the states of the system driven by that Hamiltonian.

The LR invariant $I(t)$ of the Hamiltonian $H_0(t)$, satisfying $\frac{dI(t)}{dt} = i\hbar \frac{\partial I(t)}{\partial t} - [H_0(t), I(t)] = 0$, is constructed as [10, 30]

$$I(t) = \frac{\hbar\Omega_0}{2}\begin{bmatrix} 0 & \cos\gamma(t)\sin\beta(t) & -i\sin\gamma(t)e^{-i\varphi} \\ \cos\gamma(t)\sin\beta(t) & 0 & \cos\gamma(t)\cos\beta(t)e^{-i\varphi} \\ i\sin\gamma(t)e^{i\varphi} & \cos\gamma(t)\cos\beta(t)e^{i\varphi} & 0 \end{bmatrix}, \quad (2)$$

where $\Omega_0$ is a constant in unit of frequency, and $\gamma(t)$ and $\beta(t)$ are time-dependent parameters to be determined [10]. More explanation on the invariant can be found in Appendix A. The invariant $I(t)$ has three eigenstates $|\phi_0(t)\rangle$ and $|\phi_\pm(t)\rangle$ with time-independent eigenvalues $\lambda_0$ and $\lambda_\pm$, and the corresponding LR phases are $\alpha_0$ and $\alpha_\pm$. The LR theory shows that the superposition of these eigenstates multiplied by the respective LR phases is a solution of the Schrödinger equation with the Hamiltonian $H_0(t)$ [31]. In this work we are interested in the eigenstate

$$|\phi_0(t)\rangle = \begin{bmatrix} \cos\gamma(t)\cos\beta(t) \\ -i\sin\gamma(t) \\ -\cos\gamma(t)\sin\beta(t)e^{i\varphi} \end{bmatrix}, \quad (3)$$

with eigenvalue $\lambda_0 = 0$ and LR phase $\alpha_0 = 0$. It is easy to check that $|\phi_0(t)\rangle$ itself satisfies the Schrödinger equation. Thus, the system Hamiltonian $H_0(t)$ can drive the qubit state evolving along the state $|\phi_0(t)\rangle$. This knowledge on the state evolution provides us the convenience of ensuring the initial and final state as we wish, in the meanwhile maintains the freedom to design the specific evolution path governed by the exact form of $\gamma(t)$ and $\beta(t)$.

The Rabi frequency $\Omega_i$ ($i = s, p$) in $H_0(t)$ relates to $\gamma(t)$ and $\beta(t)$ in the invariant $I(t)$, as follows

$$\Omega_p = 2[\dot{\beta}\cot\gamma(t)\sin\beta(t) + \dot{\gamma}\cos\beta(t)], \quad (4)$$

$$\Omega_s = 2[\dot{\beta}\cot\gamma(t)\cos\beta(t) - \dot{\gamma}\sin\beta(t)]. \quad (5)$$

Equation (4) and (5) tells that once the time-dependent $\gamma(t)$ and $\beta(t)$ are designed, the Rabi frequencies $\Omega_i$ are known. As an illustration, we consider the Hamiltonian in equation (1) drives the qubit state from the initial state $|1\rangle$ to a target state $|\psi_{target}\rangle = \cos\theta|1\rangle + \sin\theta e^{i\varphi}|0\rangle$ ($\theta$ and $\varphi$ are arbitrary angles), along the eigenstate $|\phi_0(t)\rangle$ within an operation time of $t_f$. This implies that $|\phi_0(0)\rangle = |1\rangle$ and $|\phi_0(t_f)\rangle = |\psi_{target}\rangle$, which imposes boundary conditions to $\gamma(t)$ and $\beta(t)$. However, the exact ansatzs of $\gamma(t)$ and $\beta(t)$ are still free to design as long as they meet the boundary conditions. This freedom might be used for optimizing the performance of the light pulses, such as achieving a robust quantum control against frequency detuning, and the variations in laser intensity.

The errors associated with the driving Hamiltonian can be divided into stochastic errors and systematic errors [40]. The first type of error refers to various small fluctuations in the Hamiltonian over time, such as the noise in the amplitude and phase of the light source. In many cases, this noise can be significantly reduced if the system is at a low temperature environment. Therefore, here we focus on the systematic errors, one source of which is the spatial inhomogeneity in the laser intensity. It can be described as a perturbation to the system Hamiltonian $H_0(t)$, and be treated in perturbation theory. Considering the variation in laser intensity, the total Hamiltonian of the system is

$$H = H_0 + H_1, \quad (6)$$

where $H_1 = \lambda H_0$, and $\lambda \in [0,1]$ represents the strength of the perturbation. The fidelity of achieving the target states in this case is $P = |\langle \psi_{target}|\psi'_{t_f}\rangle|^2 \approx 1 - \lambda^2 \left|\int_0^{t_f} e^{-i\alpha_+(t)}(\dot{\beta}\cos\gamma + i\dot{\gamma})dt\right|^2$, where $|\psi'_{t_f}\rangle$ is the final state attained with the perturbation theory, and we omit the terms higher than $\lambda^2$ as an approximation. To investigate how sensitive $P$ is to $\lambda$, the systematic error sensitivity is used [40], which reads

$$q_s = -\frac{1}{2}\frac{\partial^2 P}{\partial \lambda^2}\bigg|_{\lambda=0} = \left|\int_0^{t_f} e^{-i\alpha_+(t)}(\dot{\beta}\cos\gamma + i\dot{\gamma})dt\right|^2, \quad (7)$$

where $\alpha_+(t) = -\int_0^t \frac{\dot{\beta}(t')}{\sin\gamma(t')}dt'$. The closer to zero $q_s$ is, the more robust the quantum control is against variations in laser intensity. Generally, the integration in equation (7) is hard to calculate analytically, depending on the exact mathematical forms of $\gamma(t)$ and $\beta(t)$. An approximation of $q_s$ in this work can be found in equation (18) in Appendix B. In what follows, we will develop the light pulses by proposing the exact ansatz of $\gamma(t)$ and $\beta(t)$ that satisfy the boundary conditions on $|\phi_0(0)\rangle$ and $|\phi_0(t_f)\rangle$, in the meanwhile ensure $q_s \approx 0$ to guarantee the insensitivity of the quantum control in respect to the variations in laser intensity.

## 3. Model and the simulation results

The boundary conditions of $\gamma(t)$ and $\beta(t)$ restricted by $|\phi_0(0)\rangle = |1\rangle$ and $|\phi_0(t_f)\rangle = |\psi_{target}\rangle$ are

$$\gamma(0) = \gamma(t_f) = \pi, \quad (8)$$
$$\beta(0) = \pi, \beta(t_f) = \pi - \theta. \quad (9)$$

Here we propose the ansatz of $\gamma(t)$, $\beta(t)$ as follows

$$\gamma(t) = \pi + \sum_{m=1}^{\infty} A_m e^{-\frac{(t-B_m t_f)^2}{(C_m t_f)^2}}, \quad (10)$$

$$\beta(t) = -\frac{\theta}{t_f}t + \frac{\theta}{\pi}\sum_{n=1}^{\infty} a_n \sin\left(\frac{n\pi t}{t_f}\right) + \pi. \quad (11)$$

where $A_m$, $B_m$, and $C_m$ denotes the amplitude, center position, and the width of each Gaussian component, respectively. They have to be carefully adjusted so that the

Table 1. Parameters of Gaussian terms in equation (10) and optimal $a_n$ values of $\beta(t)$ in equation (11)

|  | The first Gaussian | The second Gaussian | The third Gaussian |
|---|---|---|---|
| Weight factor $A_m$ | 0.08 | 0.04 | 0.03 |
| Pulse center $B_m$ | 0.5 | 0.5 | 0.5 |
| Pulse width $C_m$ | 0.4 | 0.31 | 0.28 |

with the $A_m$, $B_m$, and $C_m$ given above, the optimized $a_n$ parameters are as follows

| $a_1$ | $a_2$ | $a_3$ | $a_4$ | $a_5$ | $a_6$ | $a_7$ | $a_8$ |
|---|---|---|---|---|---|---|---|
| 0.36 | 0.8378 | 0.04 | -0.0329 | -0.02 | -0.0639 | -0.0543 | -0.0201 |

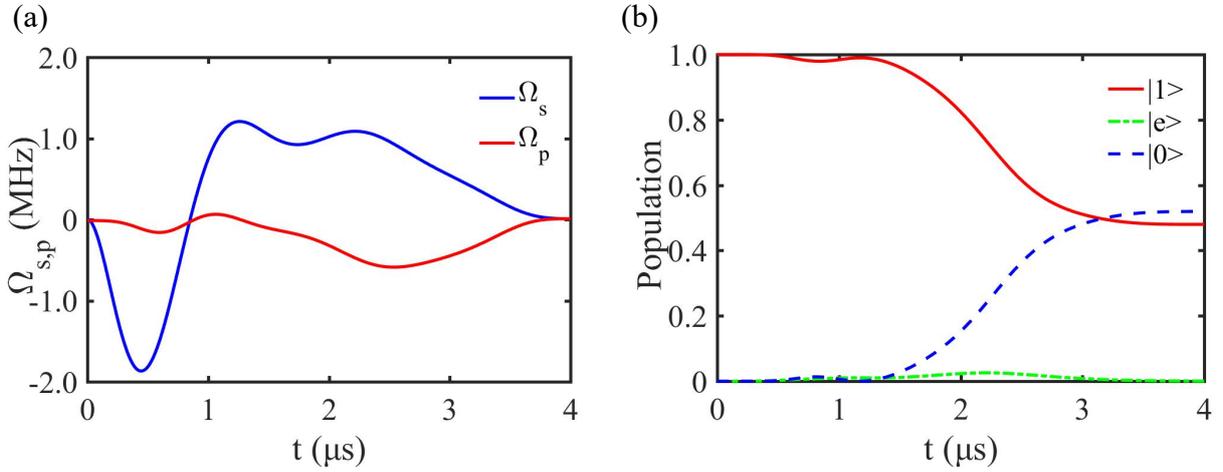

Figure 2. (Color online) (a) Rabi frequency $\Omega_s$ (blue) and $\Omega_p$ (red) of the pulses in this work. (b) Time evolution of the qubit state in level $|1\rangle$ (solid red), $|e\rangle$ (dash-dotted green), and $|0\rangle$ (dashed blue). The initial state is $|1\rangle$ and the target state is $|\psi_{target}\rangle = \frac{1}{\sqrt{2}}(|1\rangle + i|0\rangle)$.

Gaussian components contribute to $\gamma(t)$ only by a relatively small constant $\varepsilon$, especially at the starting and ending time. $a_n$ in equation (11) denotes the amplitude of each sinusoidal component, and is independent on the boundary values of $\beta(t)$.

Besides the limitation of boundary conditions and error sensitivity, Rabi frequencies are preferred to start from and end with zero in most experimental implementations, to avoid sharp changes in light intensity. The reason is that the redundant frequency components of the sharp changes could disturb the qubit operations. This demands that

$$\Omega_s(t=0,t_f) = \Omega_p(t=0,t_f) = 0. \quad (12)$$

For this, $a_n$, $A_m$, $B_m$ and $C_m$ have to satisfy

$$a_1 + 3a_3 + 5a_5 + 7a_7 + \ldots = 0, \quad (13)$$
$$a_2 + 2a_4 + 3a_6 + 4a_8 + \ldots = 0.5, \quad (14)$$

and

$$\sum_{m=1}^{\infty} A_m \cdot \frac{2(t - B_m t_f)}{C_m^2 t_f} e^{-\frac{(t-B_m t_f)^2}{(C_m t_f)^2}} \Bigg|_{t=0,\, t_f} \approx 0. \quad (15)$$

In the following passage, we will apply the pulse-designing protocol shown above to the ensemble REI qubit system to illustrate how to tailor the pulses to achieve the robust quantum operations through optimizing the multiple degrees of freedom in $a_n$. High fidelity operation in this system requires that the pulses not only interact with the ensemble qubit ions as efficient as possible regardless of the ± 170 kHz frequency detuning between them, but also address the neighboring ions which is about ± 3.5 MHz away from the qubit ions as little as possible, to suppress the off-resonant excitations. Performance of the pulses is evaluated by the operational fidelity as

$$F = \left|\left\langle \psi_{target} \middle| \psi_{t_f} \right\rangle\right|^2, \tag{16}$$

where $\left|\psi_{t_f}\right\rangle = [C_1(t_f), C_e(t_f), C_0(t_f)]^T$ is the final state of the quantum system driven by the Hamiltonian $H_0$ [10], and $C_n$ ($n = 1, e, 0$) is the probability amplitude being in level $|1\rangle$, $|e\rangle$, $|0\rangle$, respectively.

Taking $|\psi_{target}\rangle = \frac{1}{\sqrt{2}}(|1\rangle + i|0\rangle)$ ($\theta = \frac{\pi}{4}, \varphi = \frac{\pi}{2}$) and $t_f = 4$ μs as an example, we found that the pulses with parameters $A_m$, $B_m$, $C_m$, and $a_n$ as shown in table 1 could create the target state in high fidelity and with high robustness against the frequency detuning and intensity variations within the qubit ions, in the meanwhile induce low off-resonant excitations to the nearby ions closely spaced in frequency. The procedures to look for the Gaussian-related parameters ($A_m$, $B_m$, $C_m$) are shown in Appendix B, and the optimization of $a_n$ is done by using the multi-objective optimization method at given values of the Gaussian parameters. For simplicity but without losing generality, $m = 3$ and $n = 8$ are considered in this work. With all these parameters, the error sensitivity in equation (7) is numerically obtained as $q_s = 0.0137$.

The time evolution of Rabi frequencies $\Omega_{s,p}$ and the qubit state are depicted in figure 2(a) and figure 2(b), respectively. All population starts from the ground state $|1\rangle$ and ends with an equal distribution in $|1\rangle$ and $|0\rangle$ as expected. The small deviation from 50% at time $t_f$ results from the excitation by the tail of the Gaussian functions in $\gamma(t)$. Within the operation, the excited state $|e\rangle$ is hardly populated, and the time that ions spend in the excited state is only 0.04 μs. It is 17 times shorter than that in the previous work [10]. This short time strongly decreases the possibility of the spontaneous decay, and ensure a higher operational fidelity.

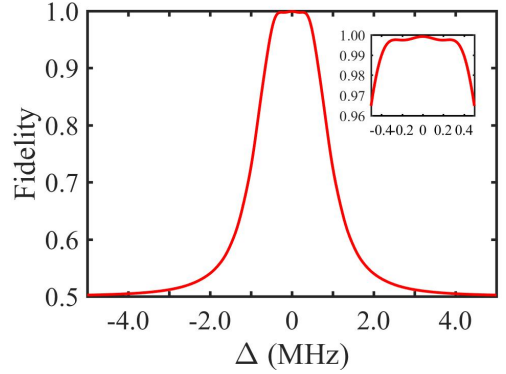

Figure 3. (Color online) Dependence of the operational fidelity on the frequency detuning, where the qubit is initially in state $|1\rangle$, and the target state is $|\psi_{target}\rangle = \frac{1}{\sqrt{2}}(|1\rangle + i|0\rangle)$.

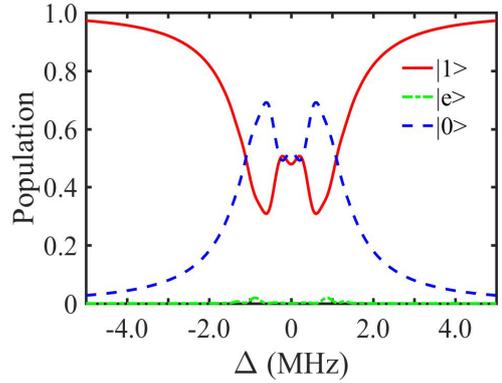

Figure 4. (Color online) Dependence of the population of the final state in level $|1\rangle$ (solid-red), $|e\rangle$ (dash-dotted green) and $|0\rangle$ (dashed blue) on the frequency detuning.

The pulses shown above can perform robust quantum control in high fidelity over the ensemble qubit in despite of the frequency detuning and variations in laser intensity. Moreover, the pulses induce low off-resonant excitations to the ions closely spaced in frequency with the qubit of interest.

*3.1 Robustness against frequency detuning and s-*

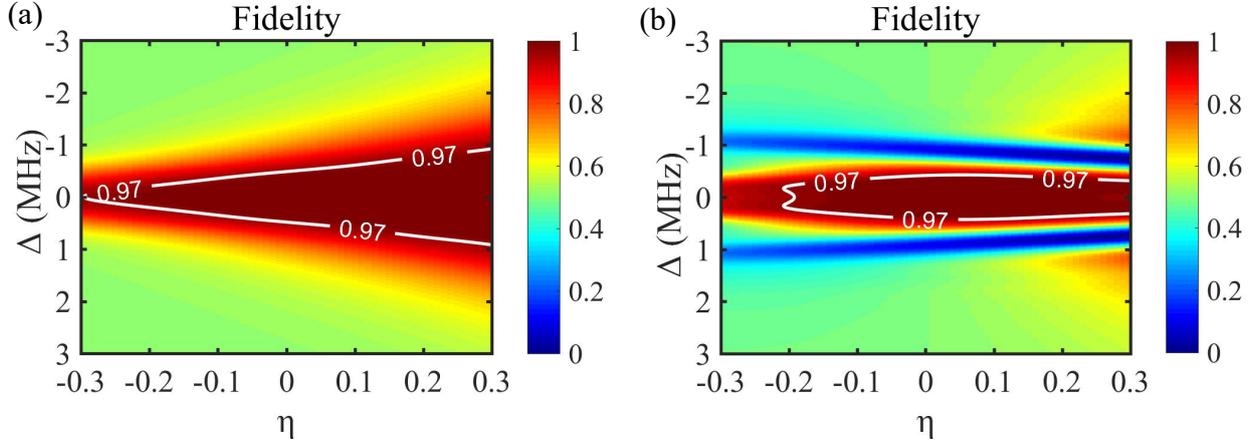

Figure 5. (Color online) Dependence of the fidelity on the fractional variation (η) in Rabi frequency at different frequency detuning (Δ). (a) using the pulses in this work. (b) using the pulses in the previous work [10].

The dependence of the fidelity $F$ on the frequency detuning is shown in figure 3. The fidelity is more than 99.7% within the frequency detuning ranges of ±270 kHz (see the inset in figure 3) which indicates that the pulses can interact with all ions within this frequency span equally well. The fidelity outside the range of ±3.5 MHz remains around 50% which is accounted for by the overlap between the initial states ( $|1\rangle$ ) and the target state ( $|\psi_{target}\rangle = \frac{1}{\sqrt{2}}(|1\rangle + i|0\rangle)$ ).

The off-resonant excitation of the neighboring ions which are closely spaced in frequency to the qubit ions were investigated by the dependence of the population in level $|1\rangle$, $|e\rangle$ and $|0\rangle$ on frequency detuning, as shown in figure 4. Within ±270 kHz around the center frequency, the population in level $|0\rangle$ and $|1\rangle$ fluctuates around 50%. Outside ±3.5 MHz, 94.2% population is in $|1\rangle$ state, which means that the off-resonant excitation is about 5.8%. It may be accepted because the ion density within this range is much less than that in the center. Furthermore, it might be reduced by sacrificing the robust region in the center via adjusting the Gaussian parameters and values of $a_n$. In addition, the ±3.5 MHz turning point could be pushed further away from the center if Eu ions with larger energy level separations were used instead of Pr ions [41].

### 3.2 Robustness against variations in laser intensity

In experiments, a Gaussian beam is widely used where the laser intensity varies with the distance to the center of the beam. This spatial inhomogeneity results in different operational fidelity at different spatial locations within the laser focus. For detecting the high-fidelity operation, a pinhole is normally used in the detection system to intentionally select the most central part of the focus, where the intensity variation is negligible, in the price of lowering down the detection signal and further the SNR. The pulses developed in this work is robust against the variation in light intensity, potentially reducing the necessity to use a pinhole and increasing the SNR.

The dependence of fidelity on the variations in the Rabi frequency at different detuning frequencies using the pulses in figure 2(a) is shown in figure 5(a), where $\eta = \delta\Omega_{p,s}/\Omega_{p,s}$ represents the fractional variation in Rabi frequency, and Δ denotes the frequency detuning. In case of zero detuning, the fidelity is above 97% over ±30% variation in the Rabi frequency. In case of non-zero detuning, the pulses are more robust against the positive variations than the negative ones. As a comparison, the same investigation using the pulses proposed in the previous work [10] was done. The results are shown in figure 5(b), where the robust range is smaller, attributed to the non-zero error sensitivity ($q_s$ = 0.5847 at $\Delta = 0$). It is 43 times larger than that with our pulses.

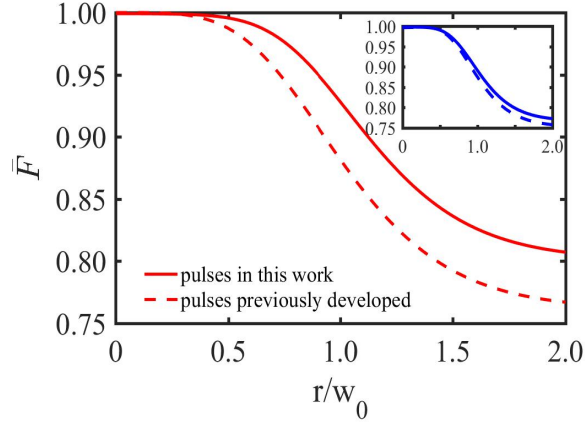

Figure 6. (Color online) The effective fidelity achieved with a Gaussian beam with a focal radius of $r$ under detection by using our pulses (solid lines) and the pulses reported in Ref. [10] (dashed lines). Red lines: on resonance. Blue lines: with a frequency detuning of 170 kHz. $w_0$ represents the beam waist.

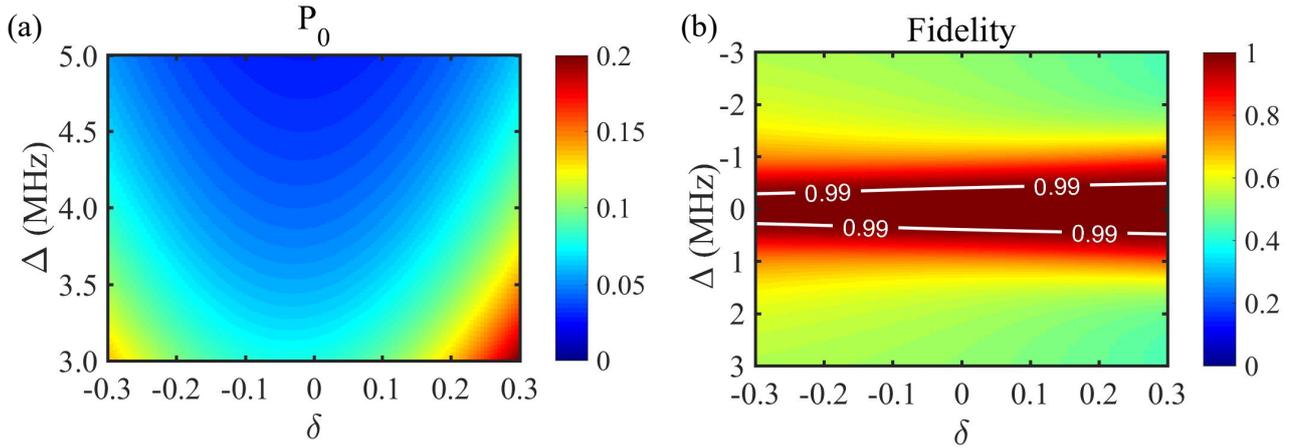

Figure 7. (Color online) Influence of the fractional variation $\delta$ in $a_2$ on (a) the off-resonant excitation and (b) the fidelity at different frequency detuning.

The effective fidelity, $\bar{F}$, achieved with a Gaussian beam with a focal radius of $r$ is investigated. The result is shown in figure 6, where the $r/w_0$ denotes the radius of the beam with respect to its waist $w_0$ at $1/e^2$. $\bar{F}$ is defined as follows

$$\bar{F} = \sum_{i=1}^{N} p(r_i) \cdot F(r_i), \quad (17)$$

where $p(r_i) = \frac{\int_{r_{i-1}}^{r_i} \Omega(r) dr}{\int_0^r \Omega(r) dr}$ ($1 \leq i \leq N, r_i = \frac{i}{N} r$) represents the weight of the fidelity $F(r_i)$ in the ring with radius from $r_{i-1}$ to $r_i$. The solid-red line (dashed-red) shows the dependence of $\bar{F}$ on the beam radius using the pulses shown in this work (in Ref. [10]) at zero detuning. The effective fidelity reaches 93% using our pulses if the signal is collected from a beam with diameter of $2w_0$, while it is only about 88% using the previous pulses. It is clear that the pulses in this work are more robust against the inhomogeneity in the light intensity benefiting from the perturbation theory. The inset in figure 6 shows the results using both types of pulses where the frequency detuning is not zero but 170 kHz. There the difference is not significant, because the perturbation theory in our framework only applies to the on-resonance interaction. The actual fidelity one would expect in experiments is a weighted combination of the red and blue curve depending on the exact profile of the absorption peaks representing the qubit.

### 3.3 Dependence of the off-resonant excitations and operational fidelity on the variation in $a_n$

In practice, light pulses experimentally generated may deviate from the optimal waveform defined by the parameters shown in table 1 limited by the electronic noises, instrumental accuracy, or variation in temperature of the waveform generators. Thus, it is interesting to investigate the dependence of the pulse performance on the variation in these parameters. As a qualitative yet instructive illustration, we considered the variation in $a_2$ since it has the largest weight among all the parameters. The results are shown in figure 7(a) and 7(b), where $\delta = \delta a_2/a_2$ is the fractional variation of $a_2$, $\Delta$ is the frequency detuning, and $P_0$ denotes the population in level $|0\rangle$. Ideally, one would expect the ions which are $\pm 3.5$ MHz away from the qubit are untouched by the pulses, that is, they remain in their initial state of $|1\rangle$. Thus $P_0 \approx 0$, as shown by figure 7(a), where only the positive axis of the detuning is plotted as the dependence is symmetric. The off-resonant excitations increase gradually with the increasing variation in $\delta$.

The operational fidelity is quite robust against the variation in $\delta$ within the frequency detuning range of $\pm 270$ kHz. The reason for this is that the population in level $|e\rangle$ hardly changes with the fluctuations in $a_2$, keeping at nearly zero. Thus, the fidelity can be simplified as $F = 0.5 + Im(C_1^* C_0)$, where the imaginary part $Im(C_1^* C_0)$ depends on the variation in $\delta$ weakly.

Above pulses are physically feasible. The magnitude of the Rabi frequency is less than 2 MHz, which can be easily achieved in the laboratory. The time-dependence profile can be initially generated by an arbitrary waveform generator as an electronic radio-frequency (RF) signal, then the RF signal drives an acousto-optical modulator to create the light pulses through the first order deflection.

## 4. Discussions and conclusions

We have proposed a method to develop robust pulses for creating an arbitrary superposition qubit state in an imperfect three-level system with high fidelity by using the invariant-based inverse engineering combined with the perturbation theory. The light pulses are built from multiple Gaussian and sinusoidal components. The multiple degrees of freedom available in the Gaussian and sinusoidal terms provides us the opportunity to tailor the pulse performance in respect to the constraints present in the quantum system. We applied this method to the REI system, and the simulation shows that the pulses developed in this work have nearly the same performance with respect to the robustness against the frequency detuning and suppression of the unwanted off-resonant excitations as the previous work [10], but outperform the latter in two aspects: (i) they are more robust against the spatial inhomogeneity in laser intensity resulting from the minimization of the systematic error sensitivity. This relieves the necessity of using a small-throughput pinhole in the detection system, thus could increase the SNR. (ii) they reduce the time that ions spend in the excited state by a factor of 17, which potentially decreases the effect of decoherence, ensuring a higher operational fidelity in a coherence-limited system. Comparing with other protocols which are insensitive to the fluctuations or imperfections in the physical system [17], the unique feature of our method is that it not only can achieve the robustness against the inhomogeneity in the laser intensity but also can tailor the interaction in the frequency domain so to achieve a nearly uniform control over a tightly packed frequency interval and suppress the interaction outside this interval. We need to say that the error sensitivity caused by the inhomogeneity in Rabi frequency was investigated in the frame work of zero frequency detuning. If the detuning is considered as well in the perturbation theory, the robustness against this inhomogeneity may be further improved, which lies out of the scope of this work.

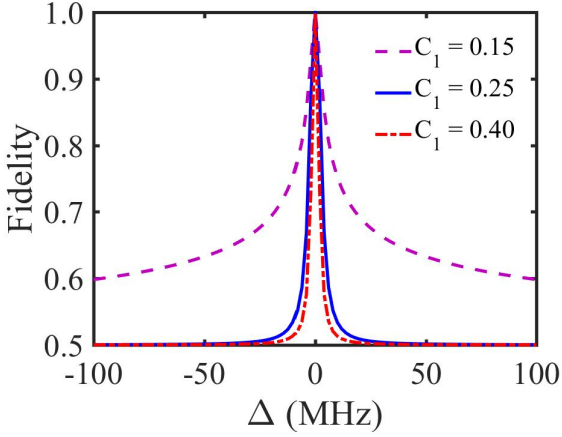

Figure 8. (Color online) Influence of $C_1$ of the single Gaussian term in $\gamma(t)$ on the fidelity. In this figure, $A_1 = 0.1$, $B_1 = 0.5$, $a_2 = 0.5$, and all the other parameters in $\gamma(t)$ and $\beta(t)$ are zeros.

The pulse-designing protocol shown in this work applies to any imperfect quantum system where qubits are closely spaced in frequency including but not limited to the REI qubit system, superconducting qubits, and the molecular qubit system. Though the protocol is introduced in a three-level system, it can be used for population transfers in a simple two-level system if only one light field is used.

**Appendix**

A. Choices of the LR invariant $I(t)$

The mathematical form of the LR invariant has many choices for a given Hamiltonian, as pointed out in the review [26]. The exact choice merely depends on the specific problems in the basis of convenience. Equation (2) in this work is inspired by the result shown in reference [30], where the invariant is constructed by utilizing the SU(2) symmetry present in the Hamiltonian. As an alternative, one may intuitively construct the invariant from its eigenstates by assigning $|\phi_0(t)\rangle$ with the appropriate form, which is favorable for the specific operation task, since the ansatz of parameters $\gamma(t)$ and $\beta(t)$ are free to choose as long as the boundary conditions are met.

B. Values of the parameters in the Gaussian functions

Here we provide a guide for choosing the appropriate values of the parameters in the Gaussian functions shown in $\gamma(t)$ as shown in equation (10).

Each Gaussian term in equation (10) are determined by three parameters, $A_m$, $B_m$, and $C_m$, which specify the amplitude, center position, and the width, respectively. Appropriate values need to be taken to ensure that the quantum control works as we expect.

There are two restrictions to the value of $A_m$. First, $A_m \neq 0$ so that the Gaussian terms contribute to suppress the off-resonant excitations caused by the sinusoidal terms in equation (11). Second, $A_m$ should be small enough so that $\gamma(t)$ deviates from $\pi$ by a small constant $\varepsilon$. With this, the error sensitivity $q_s$ is as follows

$$q_s = \left|(-i)\cdot\varepsilon\cdot\left(e^{i\frac{6\theta}{6\varepsilon-\varepsilon^3}}-1\right)\right|^2$$
$$= 2\varepsilon^2\left[1-\cos\left(\frac{6\theta}{6\varepsilon-\varepsilon^3}\right)\right]. \quad (18)$$

Clearly, $q_s$ approaches zero as $\varepsilon$ is sufficiently small. Therefore, robustness against the variations in Rabi frequency can be achieved.

For $B_m$, equation (15) needs to be evaluated, which is as follows at $t = 0$ and $t = t_f$, respectively

$$\sum_{m=1}^{\infty} A_m \frac{2(-B_m)}{C_m^2} e^{-\left(\frac{B_m}{C_m}\right)^2} = 0, \quad (19)$$

$$\sum_{m=1}^{\infty} A_m \frac{2(1-B_m)}{C_m^2} e^{-\left(\frac{1-B_m}{C_m}\right)^2} = 0 \quad (20)$$

Obviously, the best choice for $B_m$ is 0.5 so that the two equations reduce to a single one.

The effect of $C_m$ is not as transparent as that for $A_m$ and $B_m$. To investigate this, we set $A_1 = 0.1$, $B_1 = 0.5$, $a_2 = 0.5$, and all other $A_m$, $B_m$, $C_m$ ($m \neq 1$) and $a_n$ ($n \neq 2$) are zeros, and check how the operational fidelity changes in response to different values of $C_1$. The results are shown in figure 8. We found that a wider Gaussian pulse results in a narrower response in fidelity to the frequency detuning. This makes sense seen from the Fourier transform perspective. However, $C_1$ can't be too large, otherwise the boundary values of $\gamma(t)$ will deviate from its expected value of $\pi$ too much. This will eventually degrade the operational fidelity.

With all above considerations in mind, we set the $A_m$, $B_m$, and $C_m$ values as shown in table 1, and did an optimization on $a_n$ in $\beta(t)$, and developed the pulses with desired performance as shown in the main text. It is worth to note that the values given in this work are just one option. Other alternatives could potentially provide the same performance.

## Acknowledgement

The work was supported by National Natural Science Foundation of China (NSFC) (61505133, 61674112, 62074107); The International Cooperation and Exchange of the National Natural Science Foundation of China NSFC-STINT (61811530020).